\documentclass[preprint,showkeys,secnumarabic,amsfonts,showpacs,amsmath,amssymb]{revtex4}
\usepackage[dvips]{color}
\usepackage{epsfig}
\usepackage{amsmath}
\usepackage{graphicx}

\begin{document}
\title{The universe dynamics in the tachyon cosmology with non-minimal coupling to matter}

\author{H. Farajollahi}
\email{hosseinf@guilan.ac.ir} \affiliation{Department of Physics,
University of Guilan, Rasht, Iran}
\author{A. Ravanpak}
\email{aravanpak@guilan.ac.ir} \affiliation{Department of Physics,
University of Guilan, Rasht, Iran}
\author{G. F. Fadakar}
\email{gfadakar@guilan.ac.ir} \affiliation{Department of Physics,
University of Guilan, Rasht, Iran}

\date{\today}

\begin{abstract}
 \noindent \hspace{0.35cm}

Recently, the tachyon cosmology has been represented as dark energy model to support the current acceleration
of the universe without phantom crossing. In this paper, we study the dynamics of the tachyon cosmology in which the field plays the role of tachyon field and also non--minimally coupled to the matter lagrangian. The model shows current universe acceleration and also phantom crossing in the future. Two cosmological tests are also performed to validate the model; the difference in the distance modulus and the model independent
Cosmological Redshift Drift (CRD) test.

\end{abstract}


\maketitle
\section{introduction}

The recent astrophysical data indicate that there is a dark energy (DE) providing approximately two third of the current universe energy density which explain the current cosmic acceleration \cite{Reiss}--\cite{Halverson}. The most obvious candidate to explain DE is the cosmological constant which can fit observations well. However it is so small ( of order $10^{-33}eV$), in comparison with the Planck scale ($10^{19}GeV?$) that suffer from fine-tuning and the coincidence problems \cite{Mohapatra}. Numerous other DE models are produced by some exotic matter like phantom field with negative energy or some other matter scalar fields to explain the universe acceleration \cite{Caldwell}--\cite{Piao}. Unfortunately, such scalar fields are usually very light and their coupling to matter should be tuned to
extremely small values in order not to be conflict with the equivalence principle. In a sense, the cosmological evolution of the scalar fields contradict with the solar system tests \cite{Nojiri}. It is most often the case that such fields interact with matter, i) directly due to a lagrangian coupling, ii) indirectly through a coupling to the Ricci scalar or as the result of quantum loop corrections \cite{Biswass,faraj}.

Among scalar field theories, the non-minimal coupling of the field to matter lagrangian and tachyonic models, separately, have been widely investigated in last few years \cite{mohseni2}--\cite{wung}. In particular, relevant works in both interacting and non interacting cases can be found in \cite{Bagla}--\cite{setare2}. In the frame of tachyon cosmology the late time acceleration of the universe is described whereas the phantom crossing does not occur in these models. Here, we extend the previous works by integrating both models in which the scalar field in the formalism plays two roles; as a tachyon field plays the role of DE, and as a scalar field coupled to the matter lagrangian interacts with the matter in the universe and intermediate between matter and DE. The model is fully capable to represent the current acceleration in the universe and phantom crossing.

To validate the model we require various observational probes
in different redshift ranges to understand the expansion history of
the universe \cite{romano}. Moreover, to understand the true nature of
the driving force of the accelerating universe, mapping of the cosmic expansion
is very crucial \cite{Linder}. Here, we examine two observational tests
in different redshift ranges to  explain the expansion of
the universe and current acceleration \cite{Nordin}--\cite{Linder2}. The first probe is the observations of the luminosity distance –- redshift relation for the observational data on Type Ia supernovae (Sne Ia) that verifies the late-time accelerated expansion of the universe. The second probe we investigate is ``Cosmological
Redshift Drift'' (CRD) test which maps the expansion of the universe
\cite{Lis}--\cite{Jain} and measures the dynamics of the
universe directly via the Hubble expansion factor.

The paper is organized as follows: In Section two, the model
is presented with a discussion on the condition for phantom crossing behavior and acceleration expansion. Two cosmological tests are performed in section three to validate the model with
experimental data. In Section four, we present the summary and remarks.

\section{The Model and its phantom crossing condition}

We consider the tachyon cosmology with non-minimal coupling to matter given by the action:
\begin{eqnarray}\label{action}
S=\int[\frac{R}{16\pi
G}-V(\phi)\sqrt{1-\dot{\phi}^{2}}+f(\phi)\mathcal{L}_{m}]\sqrt{-g}d^{4}x,
\end{eqnarray}
where $R$ is Ricci scalar, $G$ is the newtonian constant gravity
and the second term in the action contains the tachyon potential $V(\phi)$. Unlike the usual Einstein-Hilbert action, the matter
lagrangian ${\cal L}_{m}$ is modified as $f(\phi){\cal L}_{m}$, where $f(\phi)$ is
an analytic function of $\phi$. This last term in lagrangian brings about the non--minimal coupled
interaction between the matter and the scalar field.
The variation of action (\ref{action})  with respect to the metric tensor components in a spatially flat FRW  cosmology yields the field equations,
\begin{eqnarray}\label{fried1}
3H^{2}&=&\rho_{m}f(\phi)+\frac{V(\phi)}{\sqrt{1-\dot{\phi}^{2}}},
\end{eqnarray}
\begin{eqnarray}\label{fried2}
2\dot{H}+3H^2&=&-\gamma\rho_{m}f(\phi)+V(\phi)\sqrt{1-\dot{\phi}^{2}},
\end{eqnarray}
where we put  $8\pi G=c=\hbar=1$ and $ H=\frac{\dot{a}}{a}$  with $a$ as the scale factor of the universe. We assume a perfect fluid with $p_{m}=\gamma\rho_{m}$.
Also variation of the action (\ref{action}) with respect to the scalar field  $\phi$ leads to the equation,
\begin{eqnarray}\label{phiequation}
\ddot{\phi}+(1-\dot{\phi}^{2})(3H\dot{\phi}+\frac{V^{'}(\phi)}{V(\phi)})=\frac{\epsilon f^{'}(\phi)}{V(\phi)}(1-\dot{\phi}^{2})^{\frac{3}{2}}\rho_{m},
\end{eqnarray}
where prime `` $^\prime$ '' indicates differentiation with respect to the scalar field $\phi$ and have $\epsilon=1-3\gamma$.
From equations (\ref{fried1})--(\ref{phiequation}), one can easily arrive at the generalized conservation equation,
\begin{eqnarray}
\dot{(\rho_{m}f(\phi))}+3H\rho_{m}(1+\gamma)f(\phi)=-\epsilon\rho_{m}\dot{f}(\phi),
\end{eqnarray}
which readily integrates to yield
\begin{eqnarray}
\rho_{m}f(\phi)=\frac{f^{-\epsilon}(\phi)}{a^{3(1+\gamma)}}.
\end{eqnarray}
From equations (\ref{fried1}) and (\ref{fried2}) and in comparison with the standard friedmann equations we identify the effective energy density and pressure, $\rho_{eff}$ and $p_{eff}$ as
\begin{eqnarray}\label{roef}
\rho_{eff}\equiv\rho_{m}f(\phi)+\frac{V(\phi)}{\sqrt{1-\dot{\phi}^{2}}},
\end{eqnarray}
\begin{eqnarray}\label{pef}
p_{eff}\equiv\gamma\rho_{m}f(\phi)-V(\phi)\sqrt{1-\dot{\phi}^{2}},
\end{eqnarray}
with the effective equation of state (EoS) for the model as $p_{eff}=\omega_{eff}\rho_{eff}$. From equation (\ref{pef}) in order for the model to exhibit phantom crossing and universe acceleration we must at least require that $p_{eff} <0$. From equations (\ref{roef}) and (\ref{pef}), we yield,
\begin{eqnarray}\label{phi2}
\omega_{eff}\rho_mf(\phi) + \frac{\omega_{eff}V(\phi)}{{\sqrt{1-\dot{\phi}^{2}}}}=
\gamma \rho_mf(\phi)-V(\phi){\sqrt{1-\dot{\phi}^{2}}}.
\end{eqnarray}
The equation (\ref{phi2}), in terms of redshift, $z$, and for $\omega_{eff}=-1$ at the redshift $z=z_{cross}$, becomes,
\begin{eqnarray}\label{phi2t}
\rho_mf(\phi) + \frac{V(\phi)}{{\sqrt{1-\phi'^2 H^2(1+z_{cross})^2}}}=
V(\phi){\sqrt{1-\phi'^2 H^2(1+z_{cross})^2}}-\gamma \rho_m f(\phi),
\end{eqnarray}
where, from now on, the prime `` $^\prime$ '' means derivative with respect to the redshift $z$. From equation (\ref{phi2t}), for $\gamma>-1$, phantom crossing can be achieved if the potential $V(\phi)$ and the scalar function $f(\phi)$ have opposite signs.

An analytical discussion of the phantom crossing and current universe acceleration in more details for the model is as follows. With the standard cosmological model, we assume that the matter present in the universe, coupled to the scalar field, is cold dark matter (CDM) ($\gamma=0$). In order to close the above system of equations, we also assume an exponential form of the scalar field $\phi$ for the potential, $V(\phi)=V_{0}e^{b\phi}$, and power law form for $f(\phi)$ as $ f(\phi)=f_{0}\phi^n$ where $n, b, V_{0}$ and $f_{0}$ are arbitrary constants. The runaway behavior considered for these functions satisfies equation (\ref{phi2t}). For $\gamma =0$ and from the effective pressure, equation (\ref{pef}), in order to observe phantom crossing and also universe acceleration, we first need to have $V(\phi)>0$ (for $p_{eff}<0$). From equation (\ref{phi2t}), this implies that $f(\phi)$ has opposite sign, $f(\phi)<0$. Alternatively, if $V(\phi)>0$ and $f(\phi)>0$, the phantom crossing never occurs since equation (\ref{phi2t}) is not valid, though the universe may undergo acceleration expansion as $p_{eff}<0$. Finally, by considering $\rho_{eff}>0$, in case of $V(\phi)<0$ irrespective of the sign of $f(\phi)$, we have  $p_{eff}>0$ and thus both phantom crossing and universe acceleration do not occur.

In the following a numerical calculation has been performed to study the phantom crossing in our model. Figure 1)a,c,e) show that for the same sign of $V(\phi)$ and $f(\phi)$ and different values of model parameters $n, b, V_{0}$ and $f_{0}$, equation (\ref{phi2t}) is not valid and therefore the phantom crossing never occurs. Alternatively, in figure 1)b,d,f), for $V(\phi)$ and $f(\phi)$ having opposite signs and again different values for model parameters, crossing the phantom line occurs twice in the future.

\begin{tabular*}{2. cm}{cc}
\includegraphics[scale=.36]{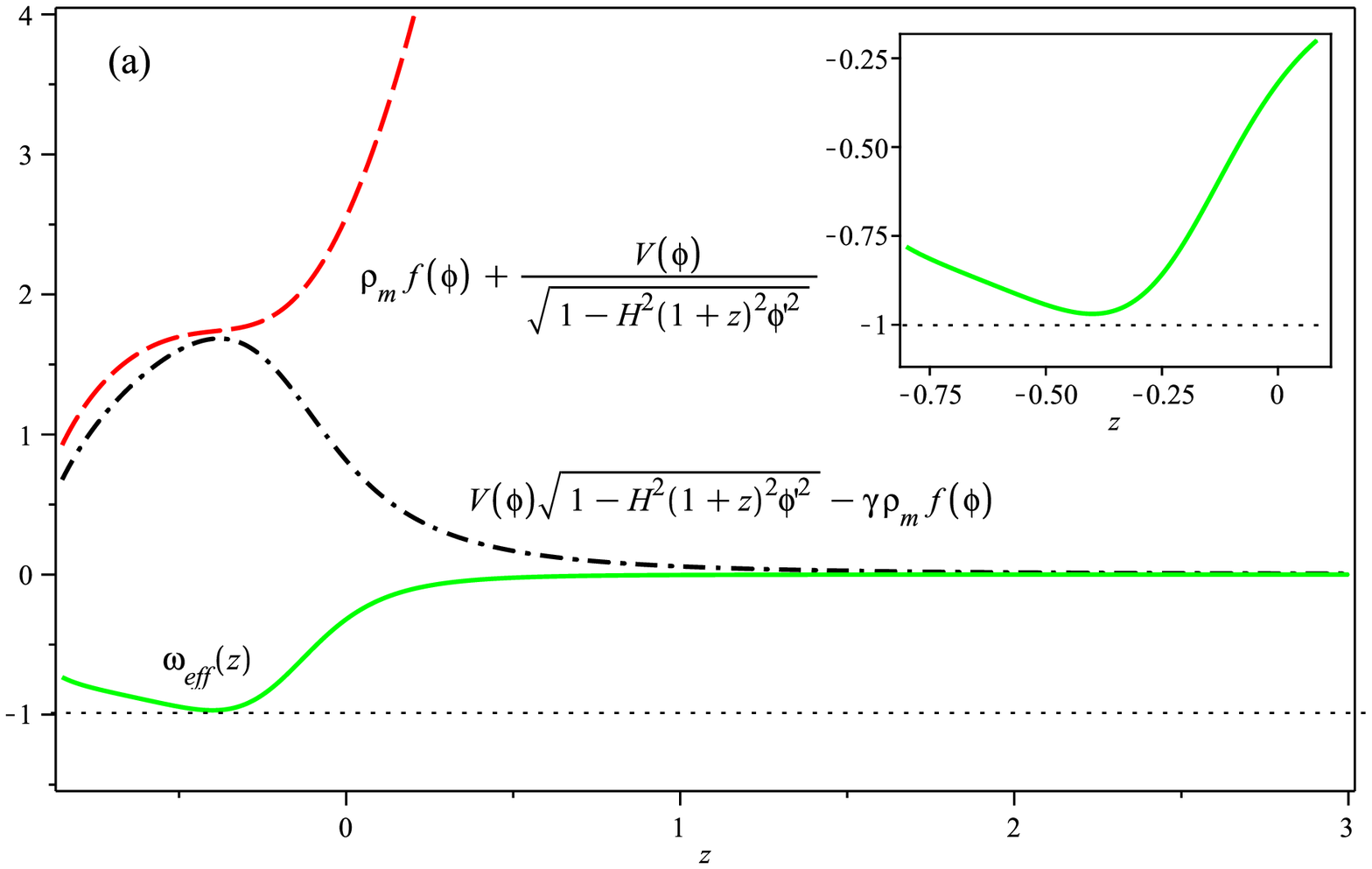}\hspace{.1 cm}\includegraphics[scale=.35]{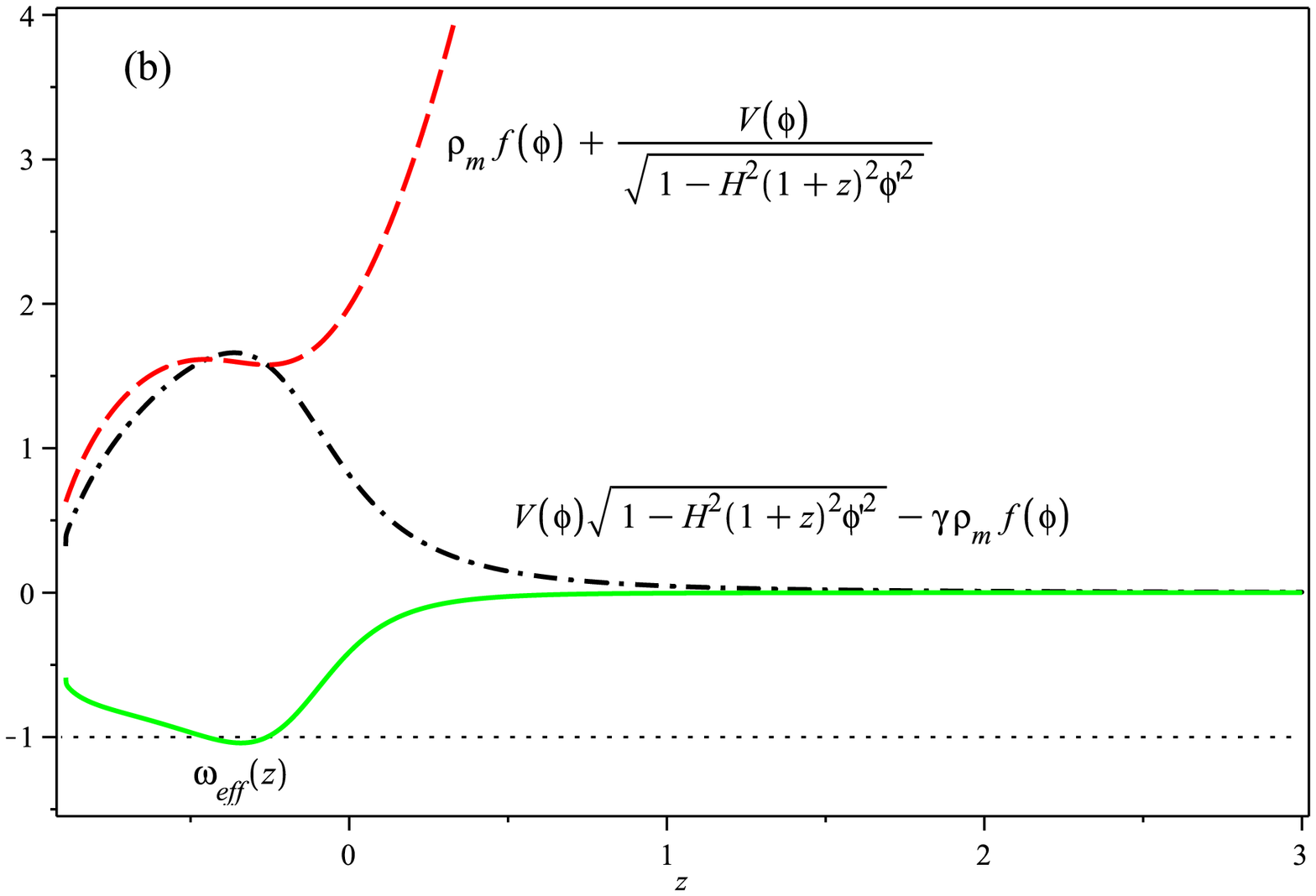}\\
\includegraphics[scale=.35]{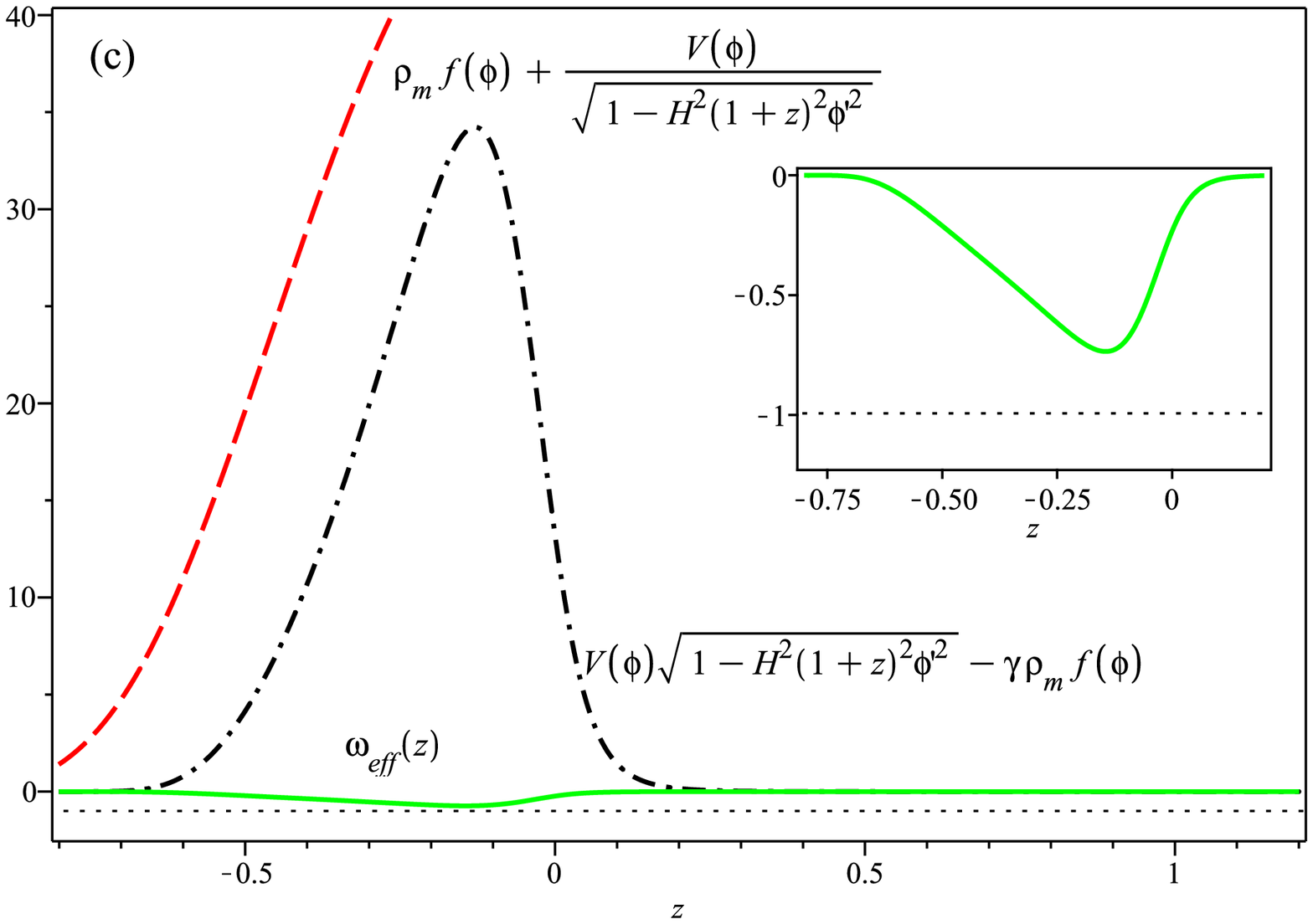}\hspace{.1 cm}\includegraphics[scale=.35]{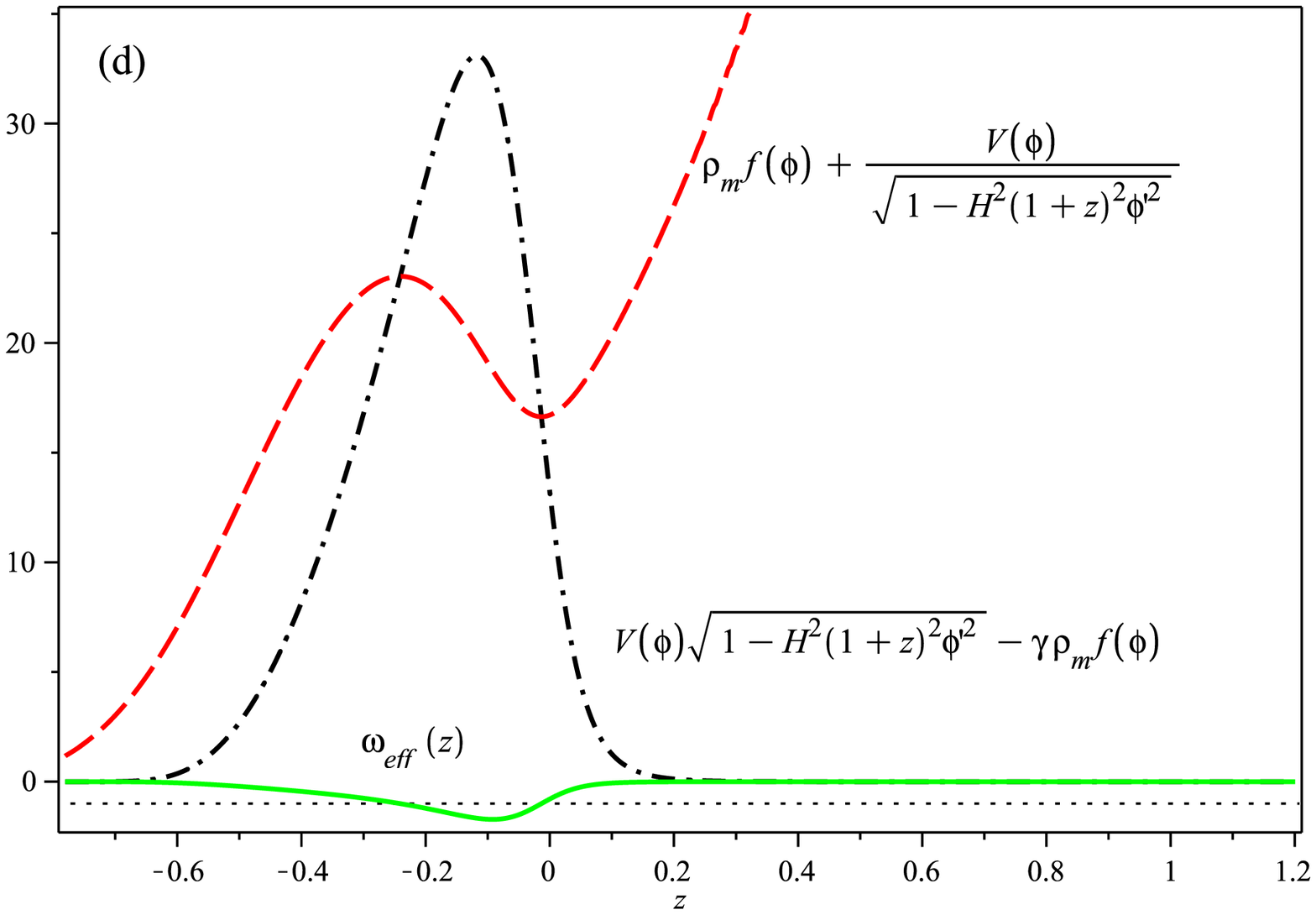}\\
\includegraphics[scale=.36]{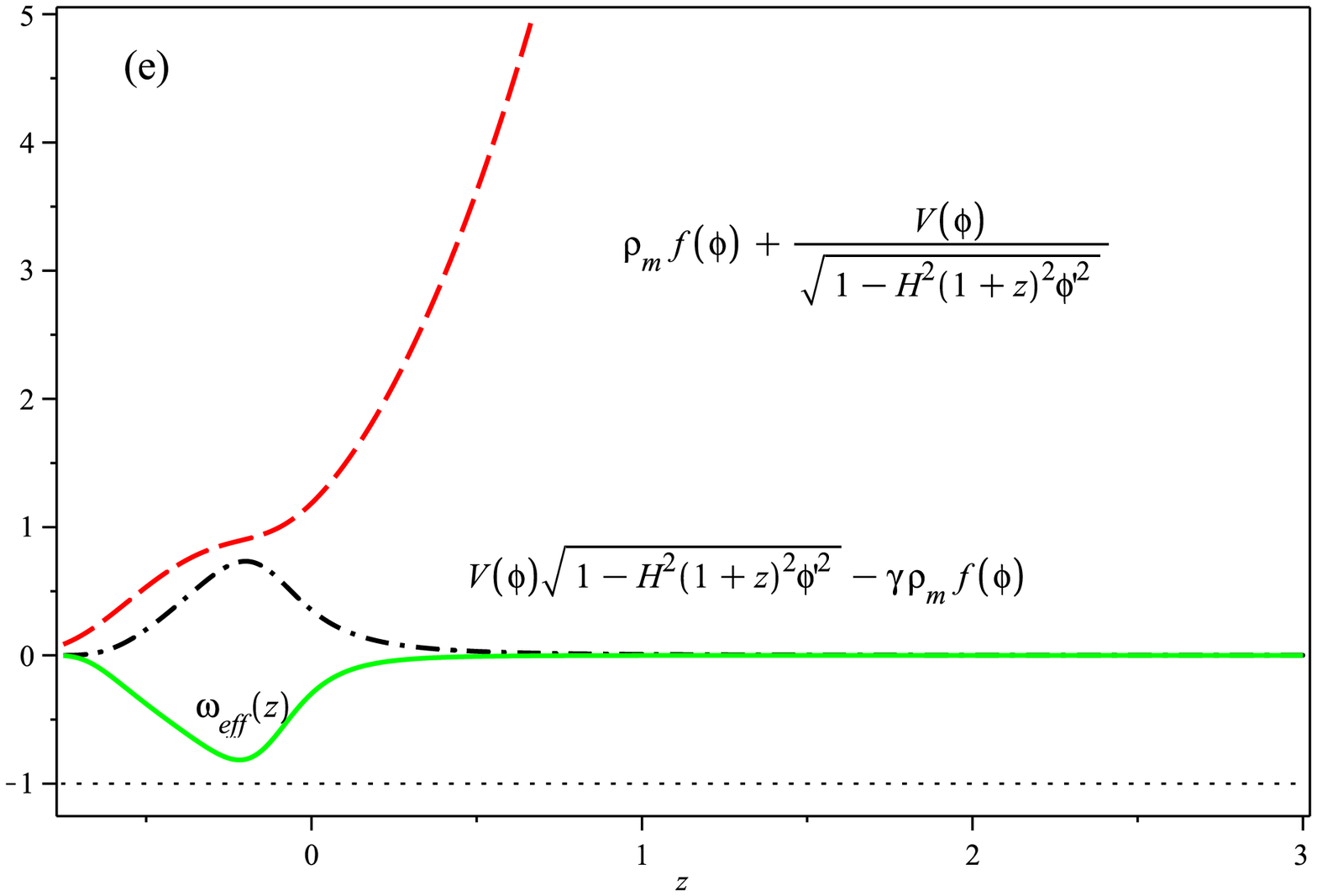}\hspace{.1 cm}\includegraphics[scale=.35]{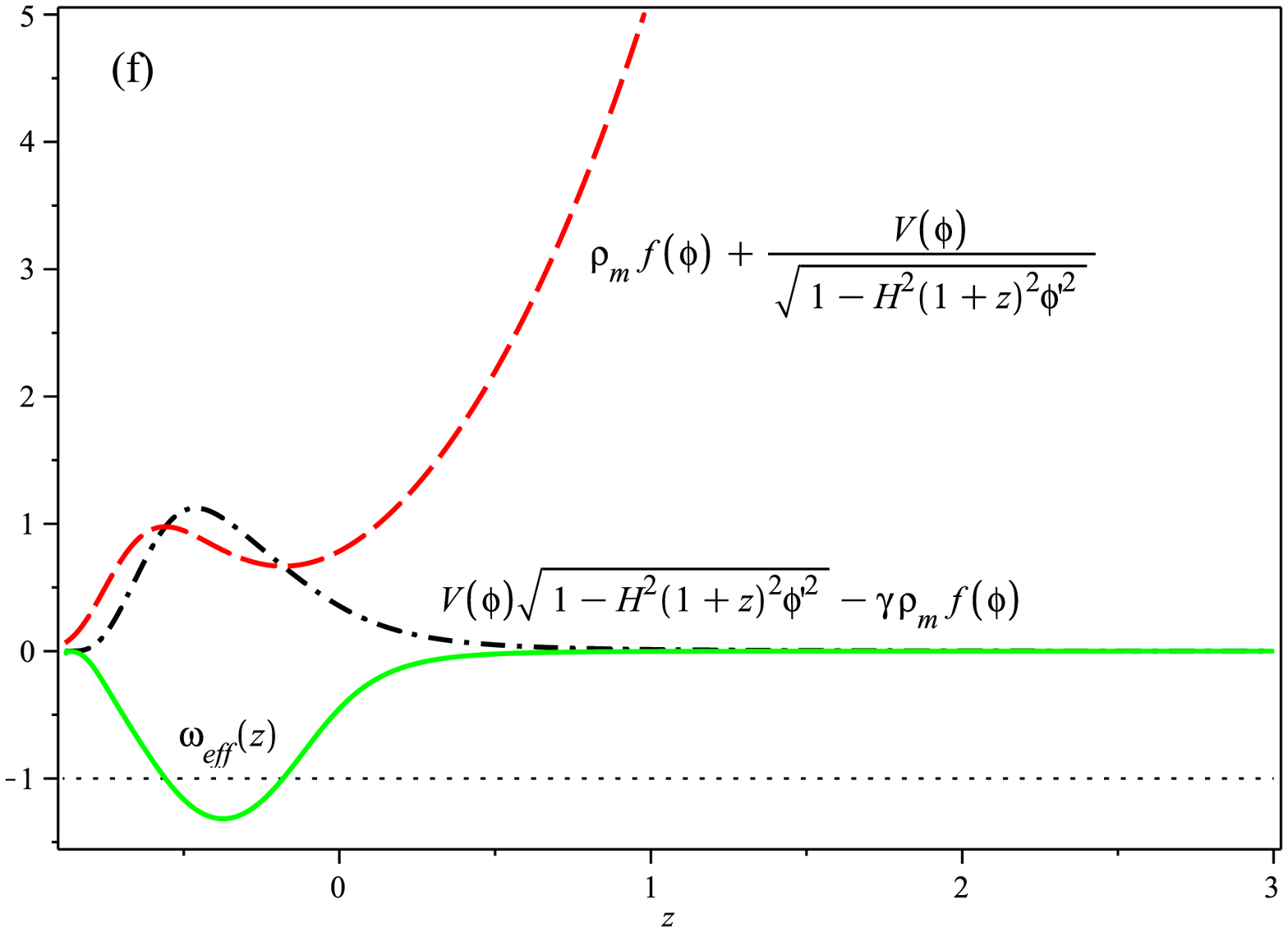}\\
Fig.1: Plots of $\omega_{eff}(z)$, $\rho_mf(\phi)+\frac{V(\phi)}{{\sqrt{1-\phi'^2 H^2(1+z)^2}}}$ and $V(\phi){\sqrt{1-\phi'^2 H^2(1+z)^2}}-\gamma \rho_mf(\phi)$\\ with $V(\phi)={\it V_{0}}\,{{\rm e}^{b\phi}
 }$ and $f(\phi)=f_{0}\phi^{n}$.\\
a) $b = 1 ,n = 1, f_0=3.5, V_0=0.5$, b) $b = 1 ,n = 1, f_0=-3.5 , V_0=0.5$. \\
c) $b = 2 ,n = -7, f_0=5, V_0=0.08$, d) $b = 2 ,n = -7, f_0=-5, V_0=0.08$. \\
e) $b = 10 ,n = 1, f_0=0.05, V_0=0.001$, f) $b = 10 ,n = 1, f_0=-0.05, V_0=0.001$. \\I.C.s$\phi(0)=1$,
$\dot{\phi}(0)=-0.8$ and $H(0)=1$.
\end{tabular*}\\

The model dependency on the initial conditions shown in Figure 2). The graphs are plotted for the case of phantom crossing for two different initial conditions. From the plots, one can see that if we further change the initial conditions, we still have the crossing behavior, but has to be magnified in a smaller proportion to be observable as in the left plot.

\begin{tabular*}{2. cm}{cc}
\includegraphics[scale=.36]{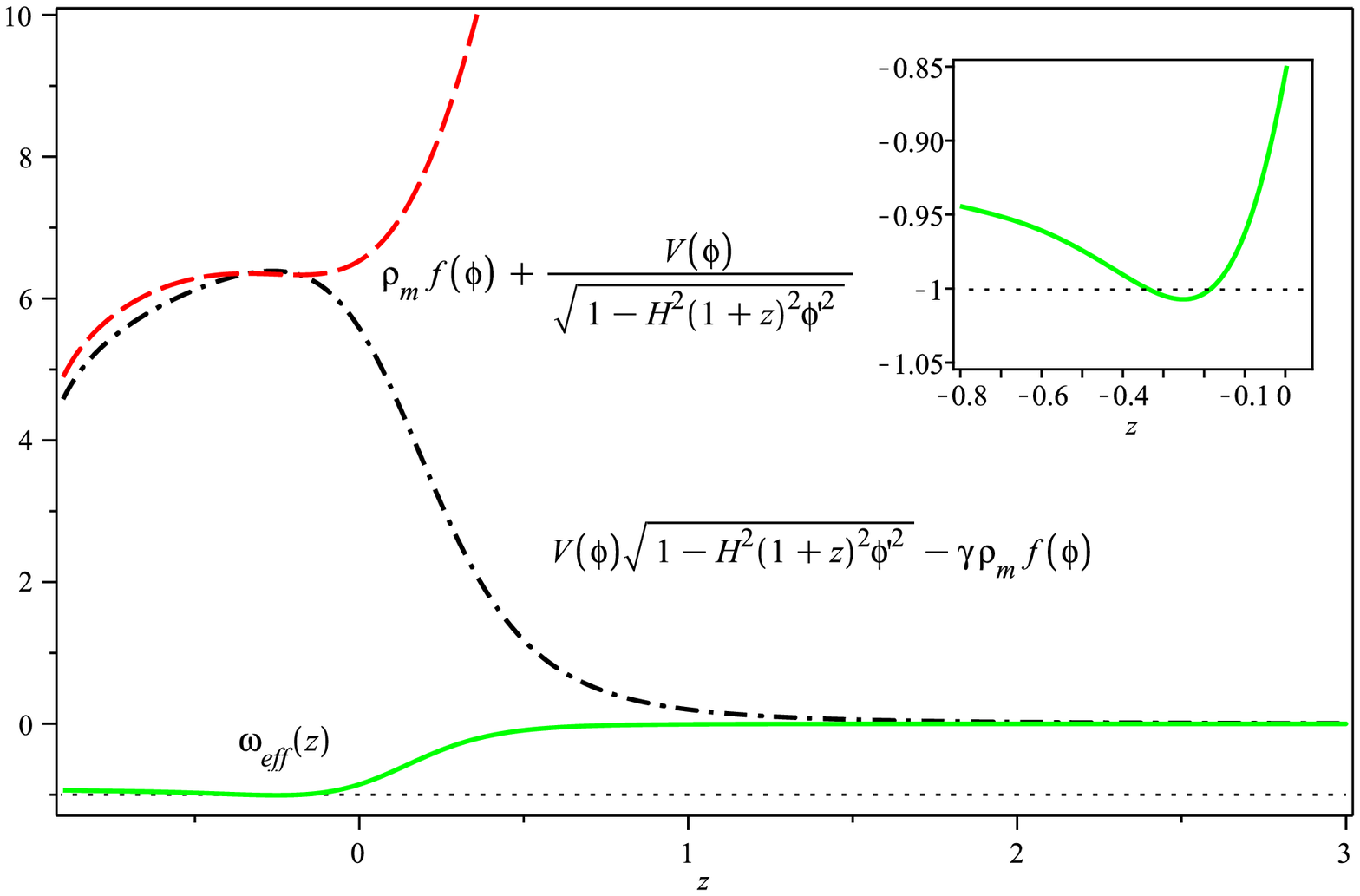}\hspace{.1 cm}\includegraphics[scale=.35]{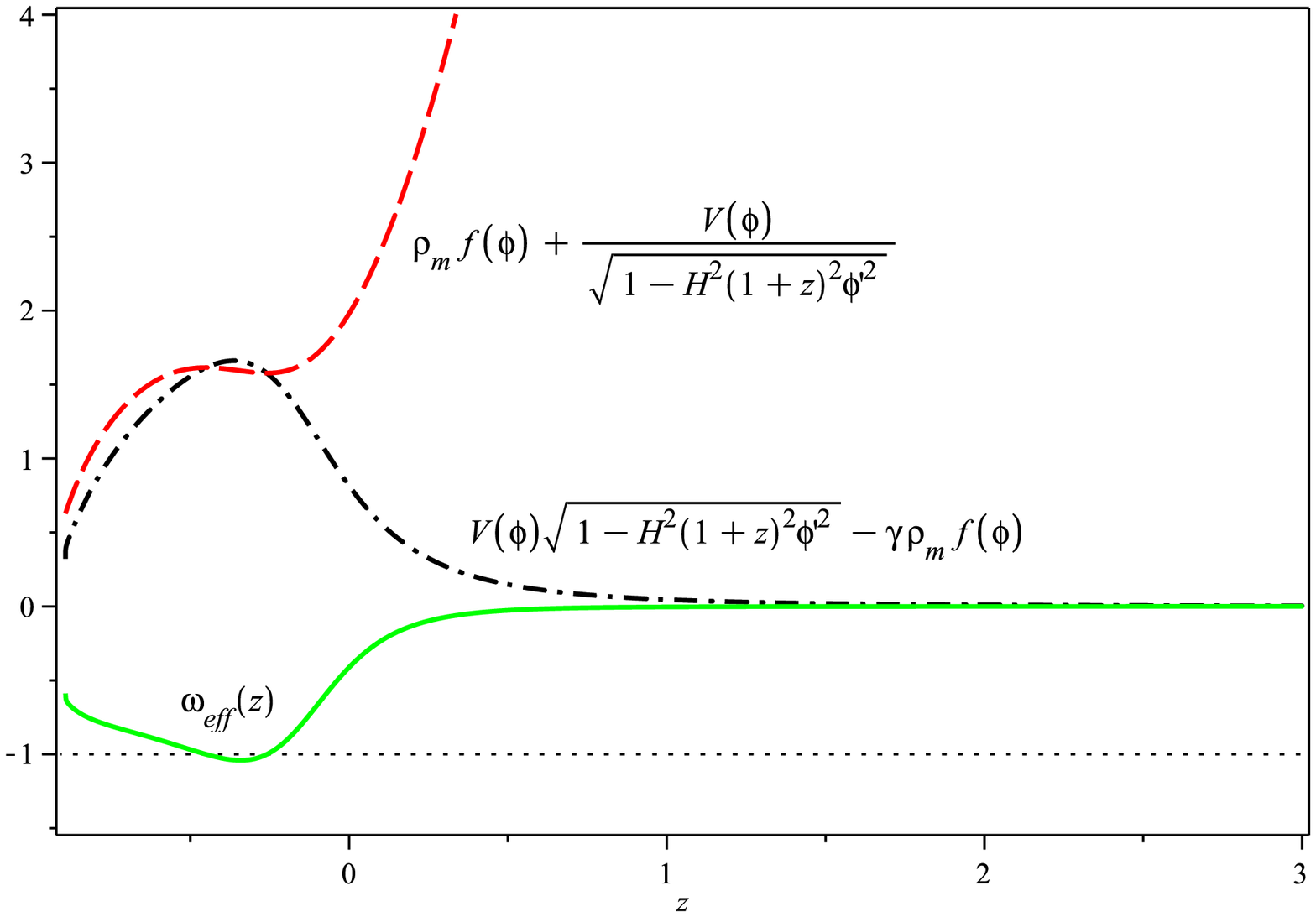}\hspace{.1 cm}\\
Fig.2: Plots of $\omega_{eff}(z)$, $\rho_mf(\phi)+\frac{V(\phi)}{{\sqrt{1-\phi'^2 H^2(1+z)^2}}}$ and $V(\phi){\sqrt{1-\phi'^2 H^2(1+z)^2}}-\gamma \rho_mf(\phi)$\\ with $V(\phi)={\it V_{0}}\,{{\rm e}^{b\phi}
 }$ and $f(\phi)=f_{0}\phi^{n}$. $b = 1 ,n = 1, f_0=-3.5, V_0=0.5$ and $H(0)=1$\\
 I.C.s: left) $\phi(0)=1$, $\dot{\phi}(0)=-0.8$ , right)  $\phi(0)=0.5$, $\dot{\phi}(0)=-0.6$   \\
\end{tabular*}\\

In the next section, in order to validate the model and support phantom crossing and current universe acceleration, we check it against observational data by performing two cosmological tests. First, our model is fitted with the distance modulus data obtained from Sne Ia data. Second, we compare our model as well as CPL (Chevallier-Polarski-Linder) parameterization model \cite{Chevalier,Linder3} with the velocity drift obtained from observational data.

\section{Cosmological tests}

There are limits from supernovae and large-scale structure data on the EoS parameter, dominated by matter and DE with the favored value near $-1$ \cite{Lis}. In this section, before we address the cosmological tests for our model, let us look more closely at the phantom crossing case and its sensitivity on the model parameters. By using $a=\frac{a_{0}}{1+z}$, its derivative as,
\begin{eqnarray}\label{hdot22}
\frac{dH(t)}{dt}=-(1+z)H(z)\frac{dH(z)}{dz},
\end{eqnarray}
and from friedmann equations (\ref{fried1}) and (\ref{fried2}), we obtain the following expression
for the effective EoS parameter,
\begin{eqnarray}\label{omegaeff}
\omega_{eff}=-1+\frac{(1+z)r'}{3r},
\end{eqnarray}
where $r= \frac{H^{2}}{H^{2}_{0}}$. In figure 3), it can be seen that for different values of $V_0$ and $f_0$, and also the model parameters $n$ and $b$, the EoS parameter crosses $-1$ at two redshift $z$ in the future, since $V(\phi)$ and $f(\phi)$ have opposite sign.

\begin{tabular*}{2. cm}{cc}
\includegraphics[scale=.45]{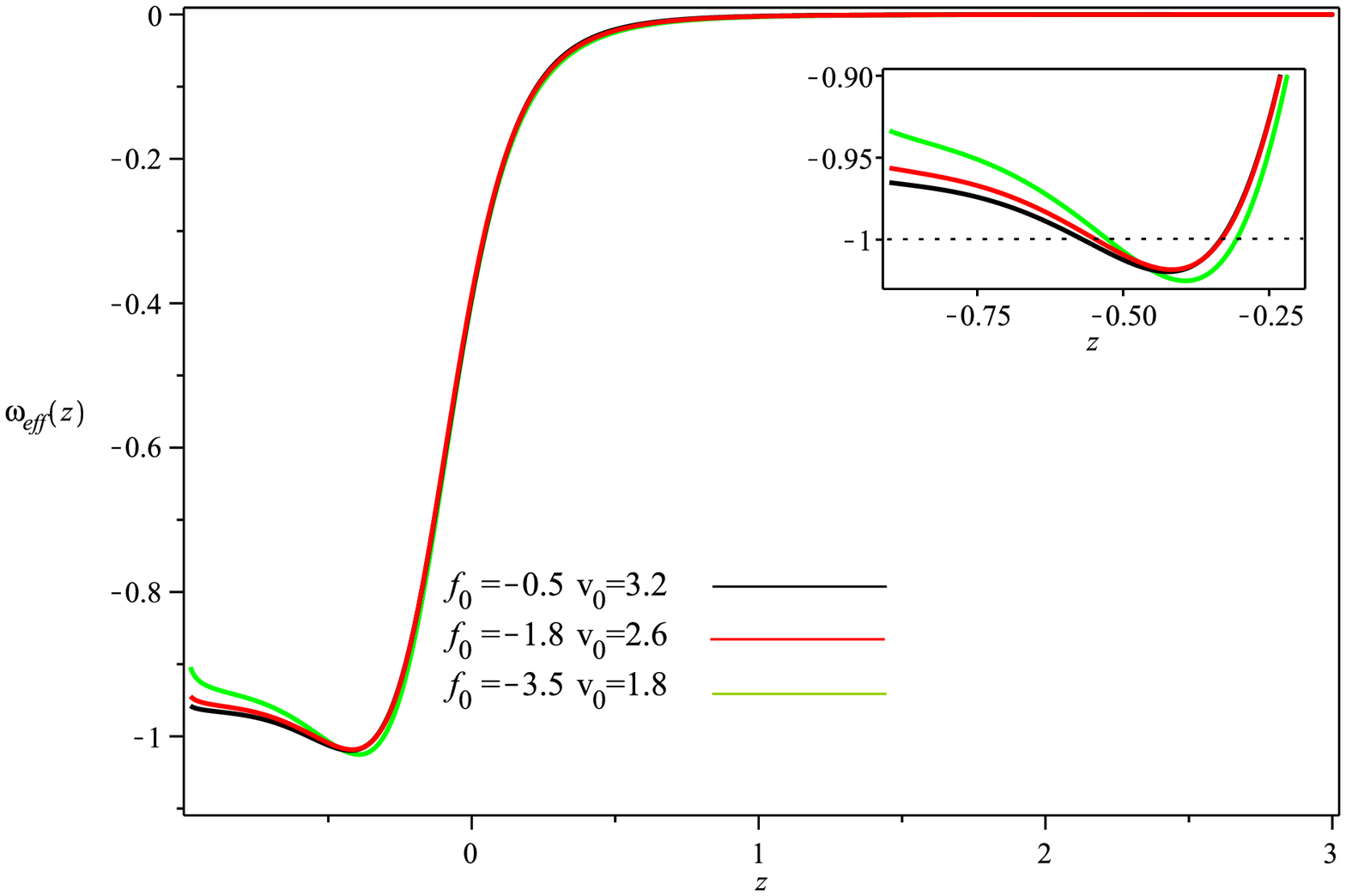}\hspace{.1 cm}\\
Fig.3: Plot of the EoS parameter, $\omega_{eff}(z)$, with $V(\phi)={\it V_{0}}\,{{\rm e}^{b\phi}}$ and $f(\phi)=f_{0}\phi^{n}$ for \\ $b = 1$ ,$n=1$. I.C.s: $\phi(0)=1$, $\dot{\phi}(0)=-0.8$ and H(0)=1.\\
\end{tabular*}

 The graph shows that for larger values of the redshift $z$ in the past, and for the given $V_0$ and $f_0$, all the effective EoS parameters approach zero as expected in matter dominated universe at high redshifts. The graph also indicates that today at $z=0$, the effective EoS parameter is about $-0.4$. Moreover, $\omega_{eff}$ will decrease before the first crossing in the future along with the decrease of the redshift $z$. The EoS dynamics implies that the behavior of DE is different in different slices of the redshifts, and inspires us to separate redshifts into several pieces and to investigate each piecewise separately. As in the previous section, the occurrence of the phantom crossing is independent of the model parameters and only depends on the sign of $V(\phi)$ and $f(\phi)$ to validate equation (\ref{phi2t}), plus $p_{eff}<0$. In high redshifts, all the trajectories approach zero and thus are independent of the values of the model parameters. However, different model parametrization can be distinguished from each other in their future behavior at redshift crossing.

\subsection{Distance modulus}

Luminosity distance quantity, $d_L(z)$, given by
\begin{equation}\label{dl}
d_{L}(z)=(1+z)\int_0^z{\frac{dz'}{H(z')}},
\end{equation}
determines DE density from observations. The difference between the absolute and
apparent luminosity of a distance object called distance modulus, $\mu(z)$, and is given by, $\mu(z) = 25 + 5\log_{10}d_L(z)$. In our model by the $H(z)$ obtained from numeric we calculate the luminosity distance and then the distance modulus. In Fig. 4, we compare our model for the distance modulus with the most recent observational data at high redshift including the Sne Ia observational data which consists of 557 data points \cite{amanullah}.\\

\begin{tabular*}{2. cm}{cc}
\includegraphics[scale=.45]{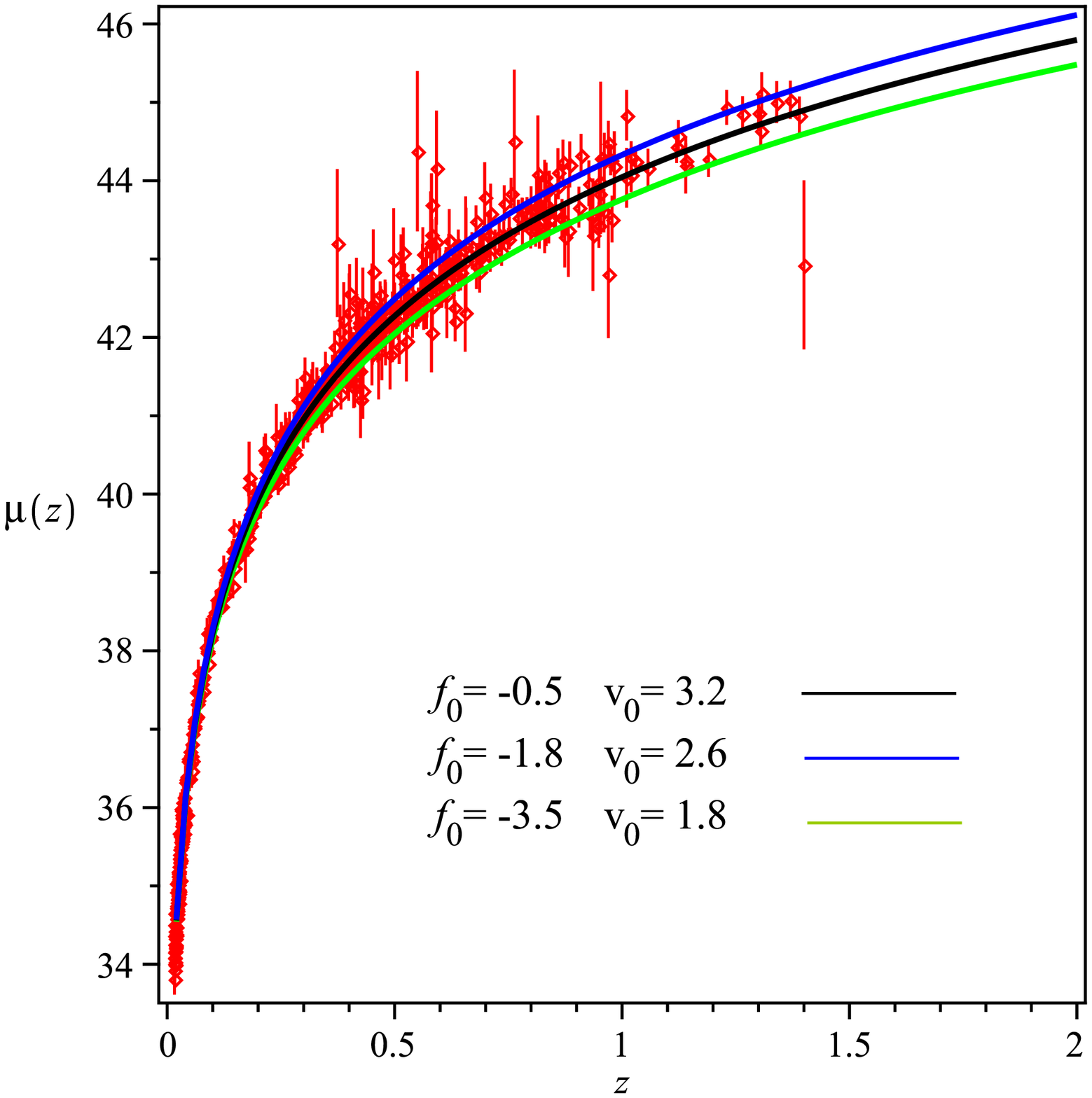}\\
Fig. 4: Plot of the distance modulus, $\mu(z)$, with respect to $z$ \\for different values of $f_0$ and $V_0$. I.C.s $\phi(0)=1$, $\dot{\phi}(0)=-0.8$ and  $H(0)=1$.
\end{tabular*}\\

The graph shows that the model with $f_0=-1.8$ and $V_0=2.6$ slightly better fits the observational data in comparison with the other cases.

\subsection{ Cosmological Redshift-Drift Test(CRD)}

Following \cite{McV}--\cite{Lake2}, the velocity drift is given by,
\begin{equation}\label{drift}
\dot{v}=\frac{cH_{0}}{1+z}[1+z-\frac{H(z)}{H_{0}}],
\end{equation}
where $\dot{v}=\frac{\Delta v}{\Delta t_{obs}}$ and $H_{0} =100 h Km/sec/Mpc $. In the
standard cosmological model $(\Lambda$CDM), the change in redshift for a time interval
of $\Delta t_{obs} = 10 yr$ is $\Delta z\simeq 10^{-9}$.
For a source at redshift $z = 3$, the corresponding velocity shift is of the order of $ \Delta v \simeq 7.5 cm/s$. To measure
this weak signal, the author in \cite{Loeb} pointed out the detection of signal of such a tiny magnitude
might be possible by observation of
the $LY\alpha$ forest in the QSO spectrum for a decade \cite{Jain}.

To observe such a tiny signals, a new generation of Extremely Large
Telescope (ELT), equipped with a
high resolution, extremely stable and ultra high precision
spectrograph is needed. Using the Cosmic Dynamics
Experiment (CODEX) operation and performing Monte Carlo simulations of quasars absorption
spectra \cite{Pasquin,Pasquini} one obtains the $\dot{z}$ measurements.

In the following, we use three sets of data (8 points) for redshift drift velocity generated by performed Monte Carlo \cite{Liske,Sur}. The data points are all for the redshift from $z=1.8$ to $z=5$ and can not validate the model with the late time acceleration of the universe. However, they can be used to support the model in comparison with  the observational data for the given range. In the following we also discuss the CPL paramterization model and compare the observational data with both CPL and our cosmological model \cite{Pasquin}--\cite{Sur}.

{\bf CPL parametrization model}

A popular parametrization which explains evolution
of DE is the CPL model \cite{Chevalier,Linder3} in which in a flat universe the time varying EoS parameter is parameterized by,
\begin{equation}
\omega_{CPL}(z)=\omega_{0}+\omega_{1}(\frac{z}{1+z}).
\end{equation}
The Hubble parameter in the model is given by,
\begin{equation}
 \Big[\frac{H(z)}{H_{0}}\Big]^{2}=\Omega_{m}(1+z)^{3}+(1-\Omega_{m})(1+z)^{3(1+\omega_{0}
 +\omega_{1})}\times\exp\Big[-3\omega_{1}(\frac{z}{1+z})\Big].
 \end{equation}
The CPL model is fitted for different values of $\omega_0$ and $\omega_1$ in correspondence with our model parametrization.

{\bf Our model}

From numerical computation, one obtains $H(z)$. Then, using equation (\ref{drift}), one leads to the velocity drift. Figure 5) shows the velocity drift against redshift $z$ for various $V_0$ and $f_0$ in both our model and CPL parametrization model. It can be seen that for $1.8<z\leq2.5$ the  black curve with $f_0=-0.5$ and $V_0=3.2$, for $2.5<z\leq4$ the red curve with $f_0=-1.8$ and $V_0=2.6$; and for $4<z<5$ the green curve with $f_0=-3.5$ and $V_0=1.8$ better fit the observational data. Furthermore, in comparison with the CPL parametrization, our model in general is in a better agreement with the data.\\

\begin{tabular*}{2. cm}{cc}
\includegraphics[scale=.6]{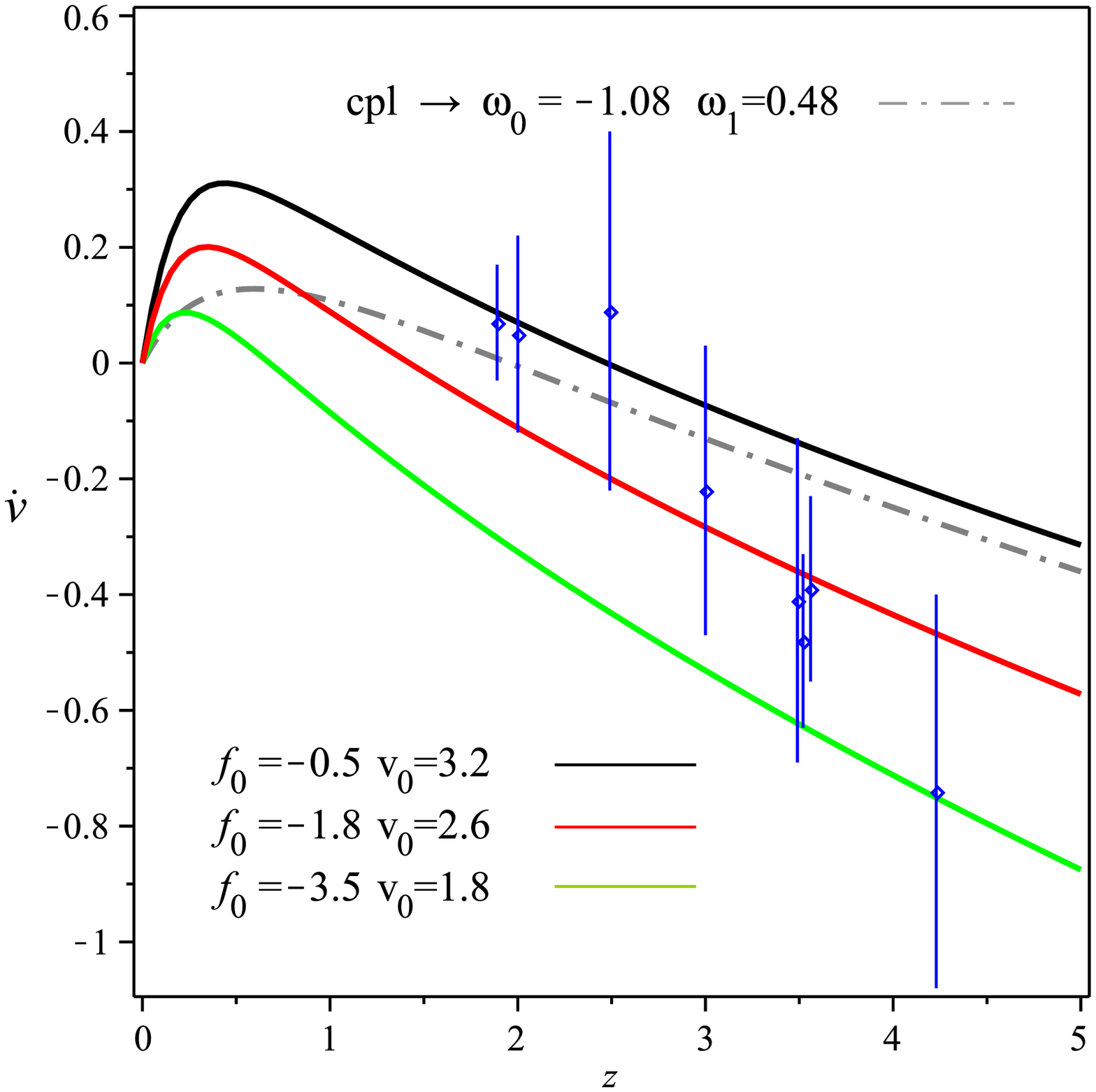}\hspace{.5 cm}\\
Fig. 5 : The velocity drift, $\dot{\nu}$, for both our model and CPL parametrization.\\
$V(\phi)={\it V_{0}}\,{{\rm e}^{b\phi}}$and $f(\phi)=f_{0}\phi^{n}$
 for different values of $f_0$ and $V_0$.($b = 1$ ,$n=1$).\\I.C.s: $\phi(0)=1$,
$\dot{\phi}(0)=-0.8$ and H(0)=1.
\end{tabular*}\\

\section{Conclusion }

In this paper, we study the dynamical evolution in a cosmological model in which the scalar field non--minimally coupled to the matter lagrangian and as a tachyon field plays the role of DE. Although in tachyon cosmology, the tachyon field cannot realize the EoS parameter crossing $-1$, in our model where the scalar field is coupled to the matter lagrangian, the phantom crossing can occur with the condition that from equation (\ref{phi2t}) the tachyon potential $V(\phi)$ and the scalar function $f(\phi)$ in the model have opposite sign. To implement this requirement, we assume an exponential behavior for the potential function and a power law form for the scalar function. For further observational tests, we also assume that the universe is filled with CDM. The model can predict the current universe acceleration and phantom crossing. The effective EoS parameter of the model, $\omega_{eff}$, transit from zero in high redshift in the past to about $-0.4$ at the present and will cross phantom divide line twice in the near future. It has been shown that the behavior of the effective EoS parameter is independent of the model parameters and only depends on the sign of the potential and scalar functions. The behavior of the parameter in the model is also independent of the initial conditions. Any changes in the initial conditions still gives us the crossing behavior if the above condition is satisfied.

We then validate the model using the observational data for the cosmological distance modulus. From the figure 4) we see that although all the trajectories with different $V_0$ and $f_0$ fit the data up to the range $z \simeq 0.2$, the trajectory with the parameter values $f_0=-1.8$ and $V_0=2.6$ better fit the data in higher redshifts. We also analyze the model with the CRD test. The variation of
velocity drift for different values of $V_0$ and $f_0$ with redshift
$z$ is shown in figure 5). A comparison between our model with the corresponding CPL parametrization shows that our model
 is in better agreement with the experimental data for different $V(\phi)$ and $f(\phi)$.

\end{document}